\documentclass[pra,aps,twocolumn,eqsecnum,showpacs]{revtex4}

\usepackage{dcolumn}
\usepackage{amsmath}
\usepackage{graphicx}
\usepackage{hyperref}
\usepackage{epstopdf}

\usepackage{ulem}

\begin{document}



\def\BE{\begin{equation}}
\def\EE{\end{equation}}
\def\BY{\begin{eqnarray}}
\def\BEA{\begin{eqnarray}}
\def\EY{\end{eqnarray}}
\def\EEA{\end{eqnarray}}
\def\L{\label}
\def\nn{\nonumber}
\def\({\left (}
\def\){\right)}
\def\<{\langle}
\def\>{\rangle}
\def\[{\left [}
\def\]{\right]}
\def\o{\overline}
\def\BA{\begin{array}}
\def\EA{\end{array}}
\def\ds{\displaystyle}
\def\c{^\prime}
\def\cc{^{\prime\prime}}

\title{High speed spatially multimode atomic memory}
\author{ T.~Golubeva, Yu.~Golubev}
\address{St.~Petersburg State University,\\ 198504 St.~Petersburg, Stary Petershof, ul. Ul'yanovskaya, 1, Russia}

\author{O. Mishina, A. Bramati, J. Laurat, E. Giacobino}
\address{Laboratoire Kastler Brossel,
Universit\'{e} Pierre et Marie Curie, Ecole Normale Sup\'{e}rieure,
CNRS, Case 74, 4 place Jussieu, 75252 Paris Cedex 05, France}
\date{\today}

\begin{abstract}
We study the coherent storage and retrieval of a very short multimode light pulse in an atomic ensemble. We consider a quantum memory process based
on the conversion of a signal pulse into a long-lived spin coherence via light matter interaction in an on-resonant $\Lambda$ -type system. In order
to study the writing and reading processes we analytically solve the partial differential equations describing the evolution of the field and of the
atomic coherence in time as well as in space. We show how to optimize the process for writing as well as for reading. If the medium length is fixed,
for each length, there is an optimal value of the pulse duration. We discuss the information capacity of this memory scheme and
 we estimate the number of transverse modes that can be
stored as a quantum hologram.
\end{abstract}

\pacs{42.50.Gy, 42.50.Ct, 32.80.Qk, 03.67.-a}%

\maketitle

\section{Introduction}
Storage and read out of quantum states of light in matter
represent a major challenge for quantum communications and quantum
information processing. Since the first proposals for such process
about a decade ago \cite{Lukin,Kozhekin2000}, and the first
experimental implementations \cite{Hau,Philips} a number of
schemes have been studied \cite{Simon,Lvovsky}. The main objective
is to store and retrieve light pulses or photons without
destroying their quantum state, that is the system should be able
to store at the same time two on commuting variables like the two
quadratures of a light pulses and allow for their retrieval.
Mapping quantum states of light like single photons or qubits
\cite{Chaneliere,Eisaman,Choi2008}, coherent light pulses without
noise added \cite{COBPG} and squeezed light \cite{Appel,Honda}
onto long lived states of atomic coherence has been first
experimentally investigated in alkali-metal gases. More recently
quantum memory for entangled photons have been demonstrated in
rare-earth doped crystals \cite{Clausen2010,Saglamyurek2010}.

While the principle of such memory registers has been mainly
developed for a single temporal and spatial mode
\cite{Gorshkov1,Gorshkov2,Gorshkov3,Gorshkov4,Hammerer2010,Drummond},
it appears that the need for multiplexing will be high and that
multimode memories need to be developed. Several schemes for a
spectral multi-mode memory have been proposed based on off
resonant Raman interaction
\cite{Mishina2007,Nunn2007,Legouet2009}, controlled reversible
inhomogeneous broadening (CRIB)
\cite{Moiseev2001,Lam,Hetet,Nunn2008} and atomic frequency comb
(AFC) \cite{Afzelius2009}, with significant recent experimental
advances in vapors \cite{Hetet1,Hosseini2010,Hedges2010,Walmsley}
and crystals \cite{Riedmatten2008}. In the direction towards
spatial multimode storage of a quantum image, a quantum hologram
scheme was proposed in reference \cite{Denis1}, based on quantum
non demolition measurement (QND) type interaction and in reference
\cite{Denis2} based on Raman-type interaction in non-collinear
field configuration. Experimentally storage of a classical image
was demonstrated in an atomic vapor based on electromagnetically
induced transparency (EIT) type interaction \cite{Shuker2008}. In
this paper, we propose a different storage protocol which
significantly combines both spectral and spatial types of
multiplexing. Our scheme provides a storage of a quantum image
carried by a very short broadband light pulse in an atomic medium.
We show that it allows both spectral and spatial multimode storage
with a good efficiency.

We consider light storage in a three-level medium in a
$\Lambda$-configuration, with a strong driving field close to
resonance with one of the transitions and a weak signal field close
to resonance with the other transition. The two lower levels are
assumed to be long-lived, being sublevels of the ground state.

The quantum signals carried by the signal pulse will be stored in the long lived  ground state coherence of the atomic ensemble. Contrary to other
schemes, based on EIT, we consider very short pulses, actually much shorter than the excited state decay time. Interaction with the medium is then
very fast and it does not allow for the build up of EIT. In contrast to echo-type memory, we consider the simultaneous action of both signal and
driving fields on the medium.

Storage based on off-resonant Raman interaction was recently proposed and demonstrated for a light pulse much broader than the natural linewidth
\cite{Nunn2007,Walmsley}. The main difference with our protocol is that we propose to explore a resonant interaction and rely on more efficient
light-mater coupling. Contrary to CRIB and AFC based schemes our protocol does not deal with an inhomogeneous broadened medium and does not use any
rephasing process. However, we will see that the quantum information storage can be quite efficient, in agreement with \cite{Gorshkov2}.

The article is organized as follows. In Sec.~\ref{model} the physical model of a 3D-memory process based on the resonant interaction of the
three-level atoms with short pulses is discussed in detail. In Sec.~\ref{write-in} the writing of a weak quantum field in space and in time is
considered. The efficiency and optimization of writing are analyzed. In Sec.~\ref{read-out} the read-out is investigated in both forward and backward
processes. The role of the diffraction and the number of stored modes are discussed at the end of this section. In App.~\ref{A} the mathematical
aspects of the problem are considered in detail.

\section{Physical model}
\begin{figure}
\includegraphics[height=3cm]{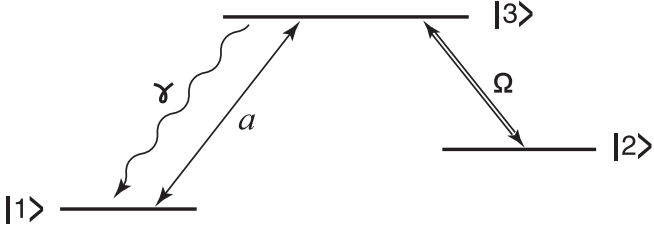}
 \caption{Three level atomic system interacting with driving field $\Omega$ and signal field $a$.}
 \label{fig1}        
\end{figure}

\L{model}
In this paper, we will study the ability of a system based on ensembles of three-level atoms (Fig.~\ref{fig1}) to store temporal as well as spatial
multimode quantum fields, thus implementing a quantum hologram. For this, we take the driving field as a plane wave, while the signal field has as a
transverse structure. A similar problem has been considered in Ref \cite{Denis1} in the case of a QND interaction of light with atoms.

The signal and driving pulses are assumed to be much shorter than the excited state lifetime $\gamma^{-1}$, so that we can neglect spontaneous
emission during the writing process, which eliminates a source of fluctuations and dissipation. For the writing  process, the driving and signal
pulses are assumed to be simultaneous and to have the same duration $T_W$. Similarly the driving pulse in the process of read-out has a duration
$T_R$ much smaller than $\gamma^{-1}$. Moreover, we assume that duration of the pulses is much larger than the time it takes for the light to go
through the atomic medium of length L
\BY
&&\gamma^{-1}\gg T_{W,R}\gg L/c.\L{duration}
\EY

The atom-field interaction will be written in the dipole approximation, and the corresponding interaction Hamiltonian
reads
\BY
&& \hat V=-\sum_j \hat d_j \hat E(\vec r_j)\L{1}.
 \EY
where $\vec r_j$ represents the spatial coordinates of the  j-th
atom and where the field $\hat E(\vec r_j)$ is a combination of two
fields, signal and driving fields, which are interacting on the two
neighboring atomic transitions
\BY
&&\hat E(\vec r,t)=\hat E_s(\vec r,t)+\hat E_d(\vec r,t).
\EY
The driving and signal fields are pulses of equal duration that verify Eq.(\ref{duration}). The driving field is in a monochromatic coherent state
with frequency $\omega_d$ and propagates as a plane wave with wavevector $k_d$. Since most of the atoms are in level $|1>$, as explained below, we
assume that we can neglect the driving field absorption and that it propagates along the z-axis through the atomic medium with a constant amplitude
$E_0$:
\BY
&&E_d(\vec r,t)=E_0e^{-i\omega_d t+i k_dz}.
\EY
We treat the signal field in the paraxial approximation as quasi-monochromatic transverse multimode wave of frequency
$\omega_s$ propagating in the same direction as the driving field with an average wavevector $k_s$. We can then write
the signal field as:
\begin{eqnarray}
\hat E_s(\vec r,t)&=&-i\sqrt{\frac{\hbar\omega_s}{2\varepsilon_0 c}}e^{-i\omega_s t+ik_sz}\hat a(z,\vec\rho,t),
\end{eqnarray}
where $\hat a(z,\vec\rho,t)$ is the annihilation operator for the
signal field and $\vec\rho=\vec\rho(x,y)$ describes the transverse
signal field distribution. The  amplitude $\hat a(z,\vec\rho,t)$ is
normalized so that the mean value $\langle\hat
a^\dag(z,\vec\rho,t)\hat a(z,\vec\rho,t)\rangle$ is the photon
number per second per unit area.

Using the rotating wave approximation we can write the Hamiltonian
as
\BY
\hat V&=&\int dz \;d^2\rho \[ i\hbar g\hat a(z,\vec\rho,t)\hat\sigma_{31}(z,\vec\rho,t)e^{\ds i k_s z}\right.
\\
&&-i\hbar g\hat a^\dag(z,\vec\rho,t)\hat\sigma_{13}(z,\vec\rho,t)e^{\ds -i k_s z}
\nn\\
&&+\left.i\hbar\Omega \hat\sigma_{32}(z,\vec\rho,t)e^{\ds i k_d z}-i\hbar\Omega^\ast
\hat\sigma_{23}(z,\vec\rho,t)e^{\ds -i k_d z}\],\nn\L{2} \EY
where $\hat\sigma_{ik}$ are the atomic coherence operators between levels $i$ and $k$ and $d_{ik}$ are the
corresponding dipole matrix elements; $\Omega=E_0 d_{23}/\hbar$ is the Rabi frequency for the driving field ; $g$ is
the coupling constant between the signal field and atom in the dipole approximation:
\BY
 && g=\(\frac{ \omega_s}{2\epsilon_0\hbar c}\)^{1/2} d_{31}.\L{3}
 \EY
For the sake of the simplicity we assume $d_{ik}$ to be real so that $g=g^\ast$ and we have set the frequency detunings $(\omega_s-\omega_{13})$ and
$(\omega_d-\omega_{23})$ equal to zero.

The collective atomic coherences are given by a sum of individual
atomic operators
 \BY
\hat\sigma_{mn}(\vec r,t)=\sum_{j} |m><n|_j \delta^3(\vec
r-\vec r_j) \nn\\
 m,n=1,2,3,\qquad m\neq n,\qquad \vec r=
\{z,\vec\rho\}.\L{4}
 \EY
These operators obey the commutation relation
 \BY
&& \[\hat\sigma_{mn}(\vec r,t),\hat\sigma_{nm}(\vec
r^\prime,t)\]=\(\hat N_m(\vec r,t)-\hat N_n(\vec
r,t)\)\;\delta^3(\vec r-\vec r^\prime), \nn\\\L{5}
 \EY
where $\hat N_m$ are the collective atomic population operators
 \BY
&& \hat N_m(\vec r,t)=\sum_{j} |m><m|_j \delta^3(\vec r-\vec
r_j).\L{6}
 \EY
In the paraxial approximation, the slow field amplitude $\hat a(\vec r,t)$  obeys the commutation relation
\BY
&&\[\hat a(\vec r,t),\hat a^\dag(\vec r^\prime,t)\]=\(1-\frac{i}{k_s}\frac{\partial}{\partial
z}-\frac{1}{2k_s^2}\frac{\partial^2}{\partial \rho^2 }\)c\delta^3 (\vec r-\vec r^\prime).\nn\\\L{8}
\EY
In the following we will assume spatially slowly varying atomic operators and make the substitution
\BY
\hat \sigma_{13} &&\rightarrow e^{\ds i k_s z} \hat
\sigma_{13},\nn\\
\hat \sigma_{12}
 &&\rightarrow e^{\ds -i (k_d-k_s) z} \hat \sigma_{12}.\L{14}
  \EY

The driving and signal fields are assumed to be superimposed at all times. They start interacting with the atomic medium at time $t = 0$. The input
plane of the atomic medium is located at $z=0$. We assume that all the $N$ atoms are initially in state $|1\rangle$.

The complete system of differential equations for the field amplitude and the collective atomic variables is given in Appendix A
(\ref{A3})-(\ref{A9}). Here, we will write the evolution equations of the system in the case where the signal field is much weaker than the control
field $|\Omega|^2\gg g^2\langle\hat a^\dag\hat a\rangle$. Then most of the atoms remain in ground state $|1\rangle$ and in the full equation
(\ref{A5}) we can replace the difference of operators $\hat N_1-\hat N_3 $ by the c-number $N$ and we can neglect the term proportional to
$\hat\sigma_{23}$ . We only keep the coherences $\hat\sigma_{12}$ and $\hat\sigma_{13}$. With these approximations, one obtain a simplified system of
the form
\BY
&&\(\frac{\partial}{\partial z}-\frac{i}{2 k_s}\frac{\partial^2}{\partial\vec\rho^2} \) \hat a(z,\vec\rho,t)=-g\;
\hat\sigma_{13}(z,\vec\rho,t),\;\;\;\;\L{9}\\
&&\frac{\partial}{\partial t}{\hat\sigma}_{13}(z,\vec\rho,t)=
 gN \hat a(z,\vec\rho,t)+\Omega
 \hat\sigma_{12}(z,\vec\rho,t),\;\;\;\;\\
&&\frac{\partial}{\partial t}{\hat\sigma}_{12}(z,\vec\rho,t)=
 -\Omega^\ast\hat\sigma_{13}(z,\vec\rho,t) .\L{11}
\EY
Due to the assumption that the light pulses are much longer than the
medium,  we have neglected the transient regime associated with the
time derivative of the signal field in Eq.~(\ref{9}).

We now take the Fourier transform of Eqs. (\ref{9})-(\ref{11}) as
\BY
\hat F(z,t;\vec q)&=&\frac{1}{2\pi}\int \hat
F(z,\vec\rho,t)e^{\ds -i \vec q\vec\rho}\;d^2\rho,\qquad \nn\\
\hat F(z,\vec \rho,t)&=&\frac{1}{2\pi} \int\hat F(z,t;\vec q)
e^{\ds i q\rho} \;d^2q ,\L{15}
 \EY
and we make the following substitutions
\BY
&& \hat a(z,t;\vec q)\to\hat a(z,t;\vec q) e^{\ds -i q^2
 z/(2k_s)},\nn\\
 &&
 \hat \sigma_{mn}(z,t;\vec q)\to\hat \sigma_{mn}(z,t;\vec q) \;e^{\ds -i q^2 z/(2k_s)}.\;\;\L{16}
 \EY

Then the set of partial differential equations giving the evolution of the system reads
\BY
&& \frac{\partial}{\partial z} \hat a(z,t;\vec q)=- g \;\hat\sigma_{13}(z,t;\vec q),\L{17}\\
 &&\frac{\partial}{\partial t}{\hat\sigma}_{13}(z,t;\vec q)=  gN \hat a(z,t;\vec q)+\Omega \hat\sigma_{12}
 (z,t;\vec q),\;\;\;\;\\
 && \frac{\partial}{\partial t}{\hat\sigma}_{12}(z,t;\vec q)= - \Omega^\ast\hat\sigma_{13}(z,t;\vec q) \L{19}.
\EY
These equation are similar to the equations obtained in Refs. \cite{Dantan,Gorshkov1,Gorshkov2} for the plane wave case. However, let us underscore
that they include the transverse spatial dependence of the signal field and they allow to treat completely the case of a multimode transverse field.
Moreover, we present here a full treatment where the hypothesis of very large values of  $ | \Omega |\gg g\sqrt{cN} $ is not made.

\section{Analysis of the writing process}\L{write-in}
\subsection{Semiclassical solutions of the equations for the writing process}\L{dimmless}

The aim of this section is to derive exact solutions for the
system of equations (\ref{17}) - (\ref{19}), in order to have a
detailed information on the efficiency of the writing and reading
processes in the atomic medium in various conditions. This will
allow optimizing the storage and retrieval processes. In order to
solve these equations, we will use the Laplace transformation in
the time domain \cite{Crisp,Gorshkov2}. In this case two of the
three differential equations are transformed into linear algebraic
equations, and the solution of the third differential equation can
be written in an explicit form. An inverse Laplace transform then
gives the values of the amplitude $\hat a(z,t;\vec q)$ and
coherences ${\hat\sigma}_{12}(z,t;\vec q)$ and
${\hat\sigma}_{13}(z,t;\vec q)$ under arbitrary initial
conditions. The detailed description can be found in Appendix A.
In the following we will concentrate on the efficiency of our
memory model, for which the semiclassical solutions are sufficient
\cite{Gorshkov1,Gorshkov2}.

According to Appendix A the semiclassical solutions for the writing stage for arbitrary relations between $g^2cN$ and $|\Omega|^2$  read
\begin{widetext}
\begin{eqnarray}
&& a^W(t,z;\vec q)=\;\int_0^t dt^\prime a_{in}(t^\prime;\vec q)\; D(t-t^\prime, z),\L{20}\\
 && \sigma^W_{12}(t,z;\vec q)= -gN\;\int_0^t dt^\prime \sin|\Omega|(t-t^\prime)\;
\int_0^{t^\prime} dt^{\prime\prime} a_{in}(t^{\prime\prime};\vec q)\; D(t^\prime-t^{\prime\prime}, z)\L{21},\\
 && \sigma^W_{13}(t,z;\vec q)= gN\;\int_0^t dt^\prime \cos|\Omega|(t-t^\prime)\;
\int_0^{t^\prime} dt^{\prime\prime} a_{in}(t^{\prime\prime};\vec q)\; D(t^\prime-t^{\prime\prime}, z),\L{22}
\end{eqnarray}
where the kernel $D(z,t)$ is expressed via the first-order Bessel
function of the first kind $J_1$ in the form
 \BY
 &&D(z,t)=\delta(t)-\cos|\Omega|t\;\sqrt{\frac{2g^2Nz}{t}}J_1\(\sqrt{2g^2Nzt}\)+\nn\\
 &&\nn\\
 &&+\frac{1}{2}g^2Nz\int_0^tdt^\prime
\[ \frac{1}{\sqrt{t^\prime}}e^{\ds -i|\Omega|t^\prime}J_1\(\sqrt{2g^2Nzt^\prime}\)\]\[ \frac{1}{\sqrt{t-t^\prime}}
e^{\ds i|\Omega|(t-t^\prime)} J_1\(\sqrt{2g^2Nz(t-t^\prime)}\)\].\L{A35}
\EY

 To examine the efficiency of the storage for each
spatial spectral field component, we assume that an amplitude of the input signal pulse is constant in time,
\BE a_{in}(t,\vec q)=a_{in}(\vec q). \L{const}\EE
In view of numerical calculations, the solutions can then be written in the following form
\BY
&& a^W(\tilde t,\tilde z;\vec q)=a_{in}(\vec q) \[1+\int_0^{\tilde t} d\tilde t^\prime \;\tilde D(\tilde t^\prime,
\tilde
z)\],\L{a_dimles}\\
 && \sigma^W_{12}(\tilde t,\tilde z;\vec q)= -p\;a_{in}(\vec q)\[1-\cos\tilde t+ \int_0^{\tilde t}d\tilde t^{\prime}
 \[1- \cos(\tilde t-\tilde t^{\prime})\]\;\tilde D(\tilde t^{\prime},\tilde z)
\], \L{sigma_dimles}\\
 && \sigma^W_{13}(\tilde t,\tilde z;\vec q)= p\;a_{in}(\vec q)\[\sin\tilde t +\int_0^{\tilde t} d\tilde t^{\prime}
 \sin(\tilde t-\tilde t^{\prime})
 \; \tilde D(\tilde t^{\prime}, \tilde z)\], \L{sigma13_dimles}
\EY
where we define a new kernel $\tilde D(\tilde t,\tilde z)$ as
\BE
D(t,z)=|\Omega|\[\delta(\tilde t)+\tilde D(\tilde t,\tilde z)\], \EE

and
\BY
&&\tilde D(\tilde t, \tilde z)=-\cos \tilde t\;\sqrt{\frac{\tilde
z}{\tilde t}}J_1(\sqrt{\tilde z\tilde t})+\frac{1}{4}\tilde
z\int_0^{\tilde t}d\tilde t^\prime
\[ \frac{1}{\sqrt{\tilde t^\prime}}e^{\ds -i\tilde t^\prime}J_1(\sqrt{\tilde z\tilde t^\prime})\]
\[ \frac{1}{\sqrt{\tilde t-\tilde t^\prime}} e^{\ds i(\tilde t-\tilde t^\prime)}
J_1(\sqrt{\tilde z(\tilde t-\tilde t^\prime)})\].
\EY
\end{widetext}
We have defined an effective interaction coefficient  $p$, given
by
$$p=\frac{gN}{|\Omega|}$$ and we have introduced dimensionless values of for time $\tilde t$
and space $\tilde z$, defined by
\BY
&&\tilde t =|\Omega| \;t,\qquad\tilde z =\frac{2g^2N}{|\Omega|}
\;z.
\EY
The above definitions can be understood in the following way. We focus our analysis on a time scale on which the spontaneous decay of the upper level
is negligible, so the effective evolution rate of the levels is determined by the Rabi frequency $\Omega$. The inverse Rabi frequency $\Omega^{-1}$
is then the natural time unit. Moreover, if we replace the relaxation constant of the upper level $\gamma$ by an effective decay rate which is the
Rabi frequency in the usual expression of the optical depth for an atomic medium of length $z$, we get an effective optical depth $2g^2Nz/|\Omega|$.
This expression is the dimensionless space coordinate $\tilde z$ defined above.

The expression for the probe field amplitude (\ref{20}) is the convolution of the field value at the input of the
medium with the kernel $ D(z, t) $. The expressions for the coherences (\ref{21})-(\ref{22}) involve double
convolutions. Rather than looking for an analytical solution, we have solved this system numerically, and in the next
sections, we analyze the results of this calculation.

\subsection{Evolution of the signal field and of the atomic coherence inside the memory cell}

As stated earlier, we study here the light-matter interface in conditions where the light pulse is much shorter than
the atomic excited state lifetime but much longer than its propagation time in the atomic medium $L/c$. Thus the
propagation time of the pulse wavefronts inside the medium is very short, and we can neglect the evolution of the field
and the atomic state and also any energy exchange between field and atoms over times on the order of $L/c$. In this
approximation, we also consider the driving field as constant during the pulse duration $T_W$. Note that the latter
approximation is not fundamentally necessary for our calculations but it simplifies the solutions.

In view of the above approximations, we can define the writing
process duration as the interval between time $ t = 0 $, when the
front part of the pulse goes out of the medium and time $ t = T_W $
when the end of the pulse reaches the input surface of the medium.

Let us normalize the amplitude $a^W(\tilde t,\tilde z,\vec q)$  of the input field with respect to its value before
it enters the medium, and  the coherence $\sigma^W_{12}(\tilde t,\tilde z,\vec q)$ with respect to its maximal
possible value at point $\tilde z=0$, which can be shown to be $-2 p a_{in}(\vec q)$ as $\tilde t=\pi$ from Eqs.
(\ref{a_dimles})-(\ref{sigma_dimles}). The normalized signal field and atomic coherence can be written as
\BY
&& a^W(\tilde t,\tilde z)=\frac{a^W(\tilde t,\tilde z,\vec q)}{a_{in}(\vec q)},\nn\\
&&\sigma^W_{12}(\tilde t,\tilde z)= -\frac{\sigma^W_{12}(\tilde t,\tilde z,\vec q)}{2 p a_{in}(\vec q)}.\L{3.12}
\EY
\begin{figure*}
\includegraphics[height=4cm]{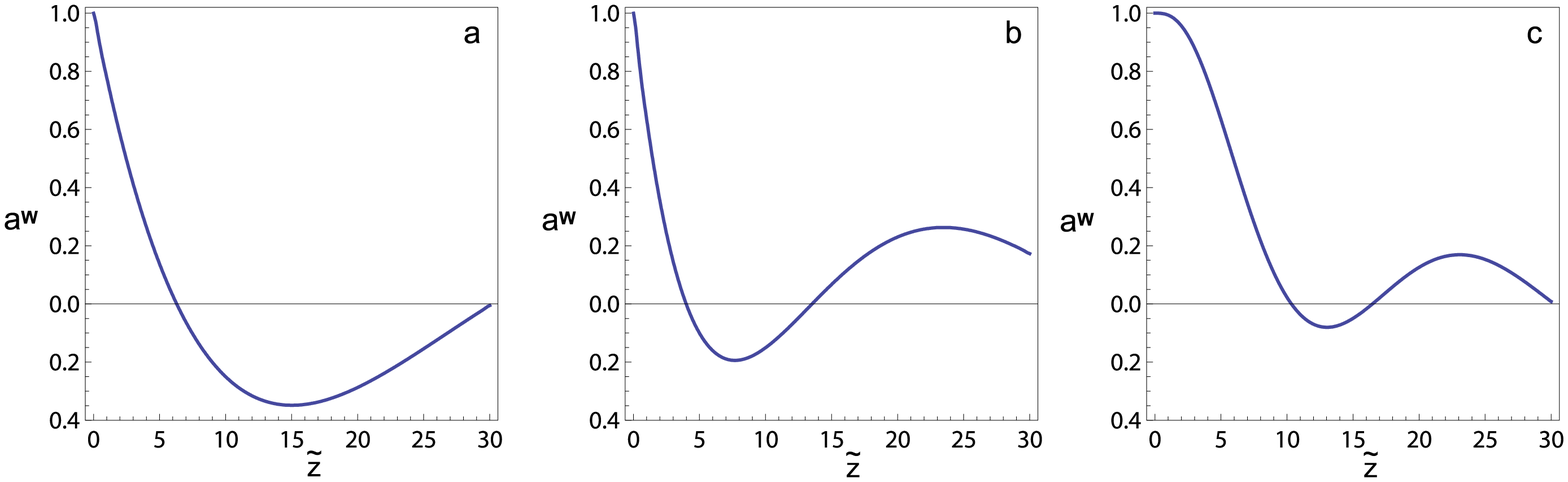}
\includegraphics[height=4cm]{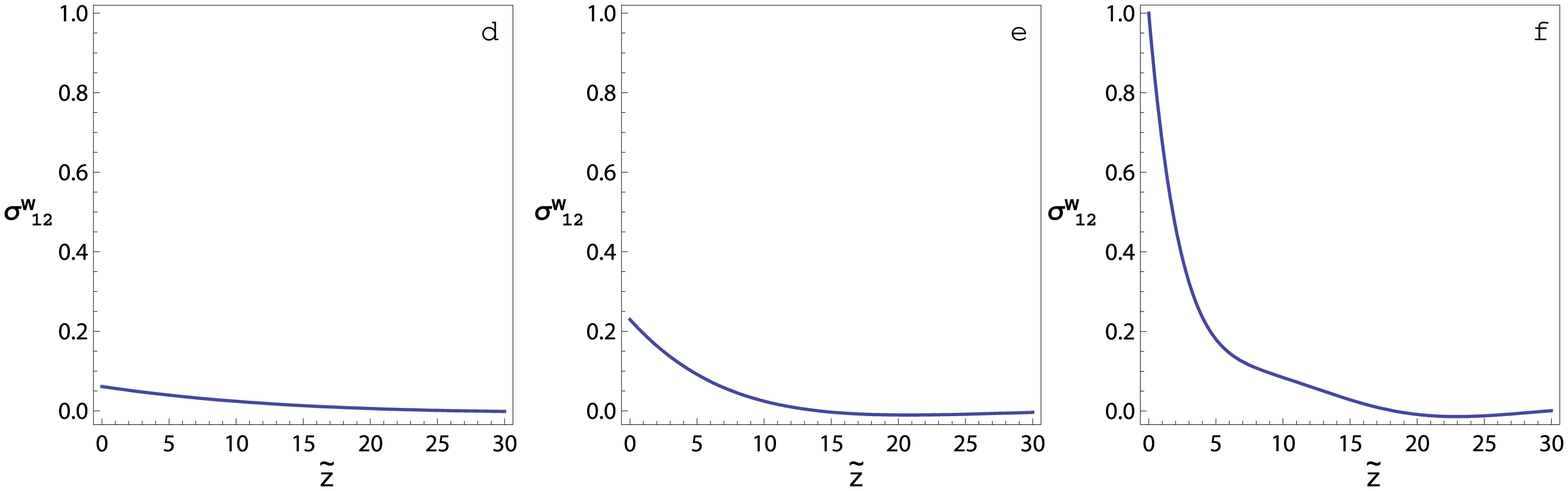}
 \caption{Normalized distributions of the field amplitude and of the atomic coherence inside the medium
 at times (a) $\tilde t=0.5$; (b) $\tilde t=1$; (c) $\tilde t=\pi$.}
 \label{fig2}        
\end{figure*}
We first analyze the distribution of the field amplitude inside the medium as a function of time. At time $\tilde t=0$ the signal field is
distributed homogeneously along the medium and is equal to 1. It remains equal to 1 at the entrance of the medium, $\tilde z=0$  till the end of the
signal pulse at $\tilde t=\tilde T_W$.

The three upper panels (a, b and c) of Fig. \ref{fig2} show the calculated distribution of field amplitude along $z$ in the atomic medium for times
$\tilde t=0.5$, $1$ and $\pi$. The amplitude of the field shows an oscillatory behaviour as a function of $\tilde z$, the effective optical depth at
point $z$. This behaviour results from the interplay between the signal field and the atoms interacting with the strong driving field.

Because of the  interaction  of the atoms with the driving and
signal fields a small fraction of the atoms initially placed in
state $|1\rangle$ is transferred to state $|2\rangle$ and an atomic
coherence $\sigma_{12}(\tilde t,\tilde z)$ is generated between
states $|1\rangle$ and $|2\rangle$. The evolution of this coherence
is shown in the three lower panels of Fig. \ref{fig2}. This
long-lived coherence is the basic feature of the quantum memory.

The lower panels in Fig. \ref{fig2} clearly show the build-up of the ground state coherence, which  results from the integrated interaction of the
atomic medium with the field since the beginning of the pulse. Our calculation allows to follow in detail the build-up of this coherence.  Up to
$\tilde t=\pi$ (which corresponds to a $\pi$ driving pulse), the coherence $\sigma^W_{12} (\tilde t, \tilde z) $ increases dramatically with time up
to its maximum value of 1 for small values of $ \tilde z$ and stays close to $ 0 $ for large values of $\tilde z$ ($ \tilde z>15 $). Thus only the
layers close to the input of the atomic medium are involved in the storage process, while deeper inside of the atomic medium, the coherence remains
equal to zero.

\begin{figure}
\includegraphics[height=6cm]{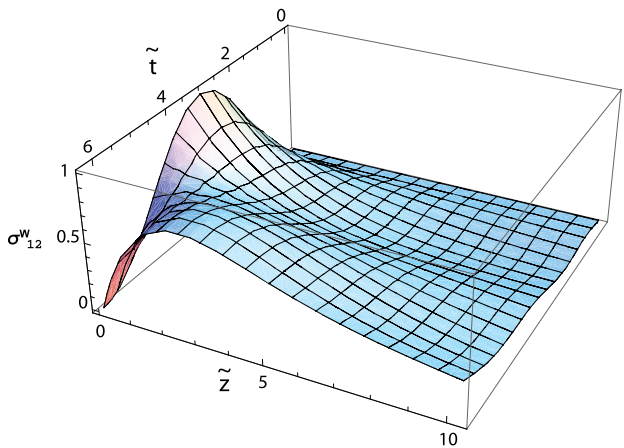}
 \caption{Distribution of the coherence $\sigma^W_{12}(\tilde t, \tilde z)$ in time and space.}
 \label{fig5}        
\end{figure}

For longer time intervals the behaviour of the atomic coherence changes. Figure \ref{fig5} depicts the evolution of
$\sigma^W_{12} (\tilde t, \tilde z)$ as a function of $\tilde t$ and $\tilde z$. At times later than $ \tilde t = \pi
$ the distribution the coherence over the medium changes: $\sigma^W_{12}$ decreases for small $\tilde z$ and starts
to grow for larger values of $\tilde z$. It can be seen clearly in Fig.~\ref{fig5} that the maximum of
$\sigma^W_{12}$ shifts from $\tilde z=0$ to non zero values of $\tilde z$.

\begin{figure*}
\includegraphics[height=4cm]{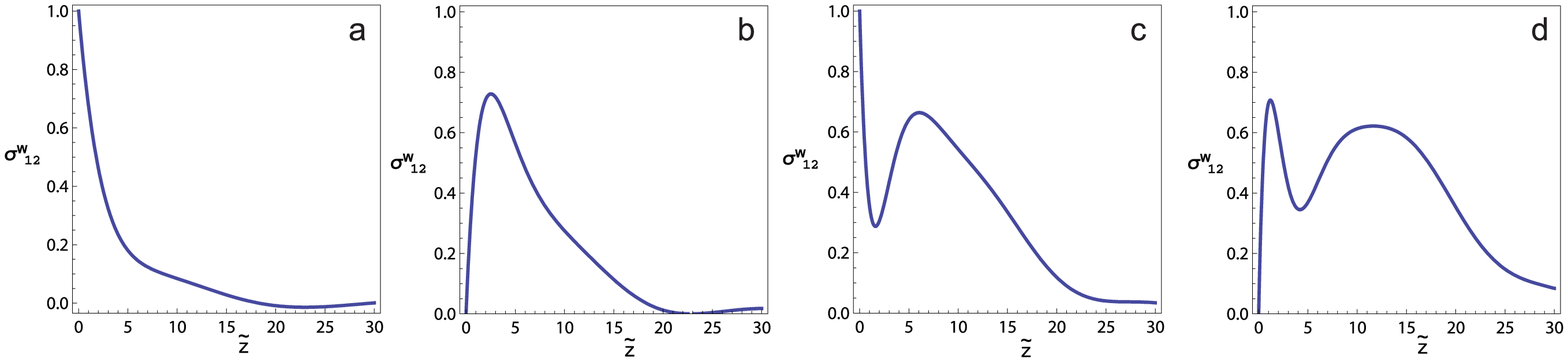}
 \caption{Distribution of the coherence $\sigma^W_{12}(\tilde t, \tilde z)$
 along the medium for (a) $\tilde t=\pi$; (b) $\tilde t=2 \pi$; (c) $\tilde t=3 \pi$; (d) $\tilde t=4 \pi$.}
 \label{fig6}        
\end{figure*}

Such a behavior is the result of two competing processes: local
writing and reading of the signal field.  Actually, after time $
\tilde t = \pi $, we can consider that the control field starts to
"read" the atomic coherence and writes it back further in the
medium. Quantum information is transferred from the input layers to
the deeper ones. Eventually, the whole medium is involved in the
storage process.

 If the duration of the field-medium interaction is increased even further,
 the maximum of the atomic coherence shifts deeper into the atomic
 medium, as shown in
Fig. \ref{fig6}, for $\tilde t=\pi$, $2\pi$, $3\pi$ and $4\pi$.

Let us note that the behavior of the coherence in a thin layer close to the input is similar to the one predicted in
Ref. \cite{Gorshkov1} for the case of a single-mode cavity: the coherence increases from 0 to 1 between times $0$ and
$ \pi $, then it decreases from 1 to 0 in the following $ \pi $ interval, etc., that is, the writing and reading
processes interchange each other with a period $\pi$. However, in our case, this is only true for $z\simeq0$. For
other points inside the medium, the behaviour is more complex than this periodic oscillation.

\subsection{Optimization of the writing process}

In the previous section, we have analyzed the writing process, that is the build-up of the ground state coherence in the atomic medium. From the
calculated distribution of $\sigma^W_{12}$, it can be seen that an efficient writing process, with an effective state exchange between signal field
and atomic coherence can only be obtained for a pulse duration larger than $ \pi$. However, as explained below, the $\pi$ pulse does not necessarily
correspond to the optimal duration because it only ensures an effective excitation of the input layers.
\begin{figure}
\includegraphics[height=4cm]{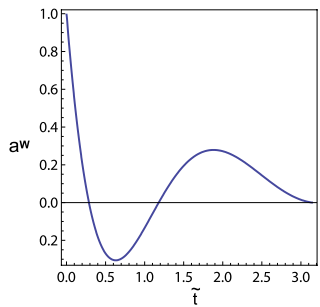}
 \caption{Field $a^W(\tilde t,\tilde L)$ at the output of the medium during the writing process for $\tilde L=10.3$.}
 \label{fig3}        
\end{figure}
In order to quantify the writing efficiency, we will use the intensity of the outgoing signal field, since any photon from the signal field going out
of the medium can be considered as a loss. Using the method presented in the previous section, we obtain the signal field at the output of the medium
of length $\tilde L$, $a^W(\tilde t,\tilde L)$. The time dependence of this field normalized according Eq. (\ref{3.12}) is shown in Fig.~\ref{fig3}
for a specific length of the medium. It can be seen that this outgoing field, usually called leakage may take non negligible values. We will
characterize the efficiency of the memory by the total losses
\BE Loss(\tilde T_W,L)=\frac{\int_0^{\tilde T_W}|a^W(\tilde t,\tilde L)|^2 d \tilde t}{\int_0^{\tilde T_W}| a_{in}(\tilde t)|^2 d \tilde
t}\times100\%\L{Loss} \EE
For a length $\tilde L$ of the medium and a duration $ \tilde T_W $
of the writing pulse, the losses correspond to the integral of the
signal field intensity over the duration of the writing pulse $
\tilde T_W $ normalized to the full pulse energy.

\begin{figure*}
\includegraphics[height=4cm]{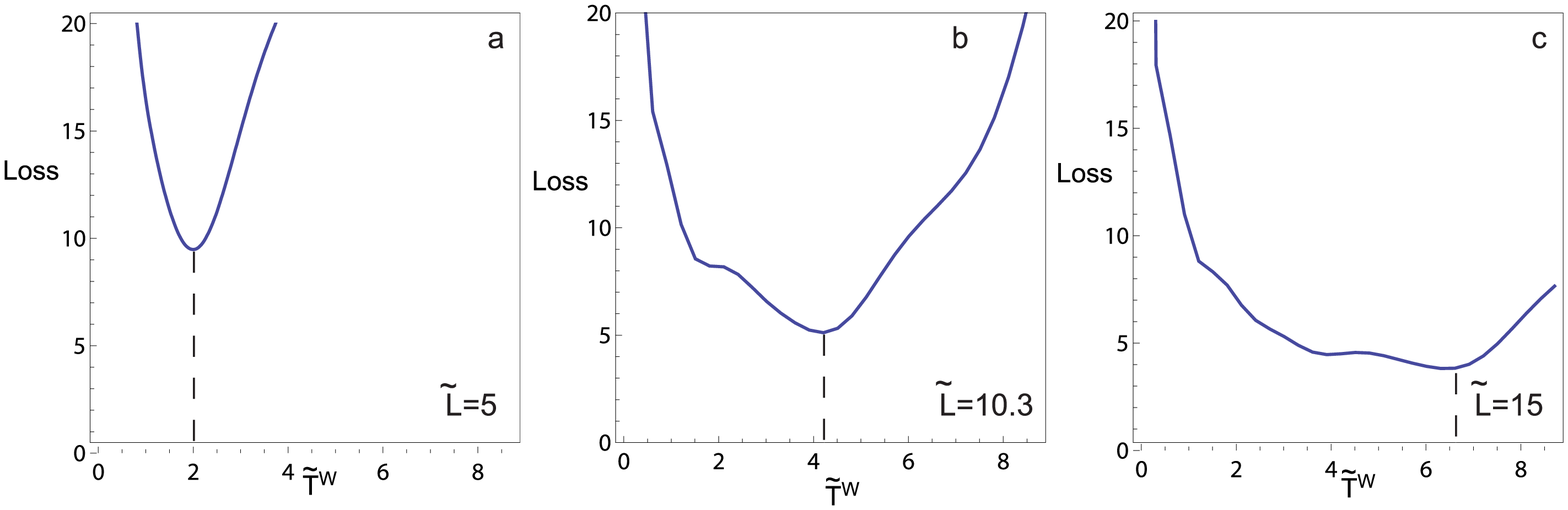}
\includegraphics[height=4cm]{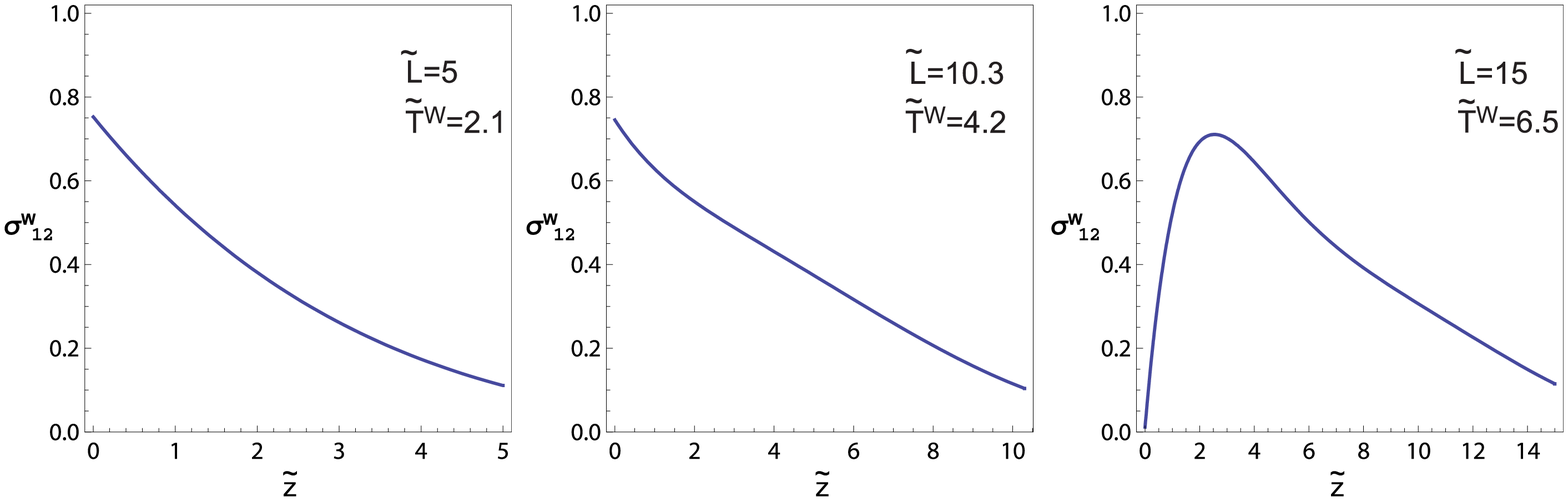}
 \caption{Writing process : relative losses in the field intensity (in percent of the input field intensity) at the output
 of the medium as a function of $\tilde T_W$ for
 (a) $\tilde L=5$; (b) $\tilde L=10.3$; (c) $\tilde L=15$ and corresponding coherence distribution for optimal conditions.}
 \label{fig4}        
\end{figure*}

Using Eq.~(\ref{Loss}) , we can optimize the writing efficiency by minimizing the losses. This is shown in Fig.
\ref{fig4}, where the upper panels depict the losses as a function of the pulse duration for various medium lengths.

Figure \ref{fig4} shows that for each length, there is an optimal
pulse duration that minimizes the losses. For a medium length of $
\Tilde L = 10.3 $, the minimal losses are obtained for $ \tilde
T_W = 4.2 $ and are about 5.1 \%. The losses decrease with longer
atomic media, but at the expense of longer writing pulses.

For each of the considered medium lengths, the lower panels of Fig.
\ref{fig4} show the distribution of the atomic ground state
coherence corresponding to the optimal writing time $ \tilde T_W $.
For short atomic media and pulse time durations, Fig. \ref{fig4}a,
losses are significant due to a field-medium interaction that does
not last long enough for the build-up of an appreciable atomic
coherence. On the other hand, if the pulse duration is longer than
optimal for a given medium length, signal field leakage takes place:
the atomic coherence starts to be re-read by the control field
resulting in a field emission that increases the losses.

\begin{figure}
\includegraphics[height=4cm]{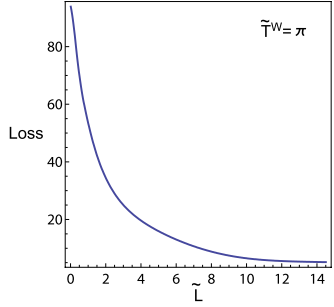}
 \caption{Writing process : relative losses in the field intensity (in percent of the input field intensity) at the output
 of the medium as a function of $\tilde L$ for $\tilde T_W=\pi$.}
 \label{fig8}        
\end{figure}

The losses $ Loss (\tilde T_W, L) $ can also be calculated as a
function of the medium length for a fixed pulse duration. This is
shown in Fig. \ref{fig8}. In this case, the losses decrease
monotonically when the medium length increases. Thus, if the
atomic medium can be considered as an unlimited resource, any
signal pulse (satisfying Eq. (\ref{duration})) can be written with
a predetermined efficiency. On the other hand, if the length of
the medium is limited, then one can no longer write an arbitrary
pulse and optimization is required.

Finally let us examine how realistic the parameters requested for optimal writing are. It can be seen in Fig. \ref{fig4}b that the ratio of $ \tilde
L / \tilde T_W $, ensuring minimum losses is approximately equal to $2.5$. Turning back to dimensional variables and separating three main factors,
we get
\BE
\frac{\tilde L}{\tilde T_W}=\frac{2g^2 N L} {\gamma}\;\frac{1}
{\gamma T_W}\;\frac{\gamma^2}{|\Omega|^2}, \EE
where $ \gamma $ is the spontaneous decay rate of the upper level. The first factor is the (real) optical depth, that is typically on the order of 1
to 10. The second factor, assuming (\ref{duration}) is much larger than 1. Thus, the Rabi frequency is determined by the inequalities
$\gamma^2/|\Omega|^2\ll 1$.  Also one can see that these optimal conditions are compatible with $g\sqrt{Nc}\sim|\Omega|$. The conditions necessary to
reach a good efficiency are expected to be quite feasible.

%


\section{Analysis of the read-out process }\L{read-out}
\subsection{Semiclassical solution of the equations for the read-out process}

The quantum information written into the atomic medium as described in the previous section can be stored during a time which is limited by the the
lifetime of the ground state atomic coherence. For times shorter than this lifetime, the quantum information can be recovered by means of the
read-out process. The latter is carried out using a driving pulse at the same frequency as the one in the writing stage. We will consider two
possible geometries for the read-out pulse propagation. The first one is the forward retrieval for which the read-out driving pulse propagates in the
same direction as the writing pulse. The second one is the backward retrieval for which the read-out pulse has the opposite direction. Like for other
memory schemes \cite{Gorshkov4,Gorshkov5,Novikova,Hammerer,Sheremet}, we will see that backward retrieval provides better read-out than forward
retrieval for a single mode while it is not necessarily true for a multimode signal.

The solutions of equations (\ref{17})-(\ref{19}) were obtained in the previous section for specific initial conditions characteristic of the writing
process. For read-out it is assumed that the signal field at the input of the medium $\hat a^R(t,z=0;\vec q)$ and the coherence at the initial time
$\hat \sigma_{13}^R(t=0,z;\vec q)$ are in the vacuum state, i.e., equal to zero in the semiclassical approach.  The ground state coherence $\hat
\sigma_{12}^R(t=0,z;\vec q)$ is assumed to coincide with its value at the end of writing stage.
First in the case of forward read-out,
 the solutions describing the reading process are given by (see Appendix A)
\begin{widetext}
\BY
&&  a^R( t,z;\vec q)=-g\int_0^z dz^\prime\int_0^t dt^\prime \sin|\Omega|t^\prime
 \sigma^W_{12}(T_W,z^\prime;\vec q)\; D(t-t^\prime, z-z^\prime),\L{36}\\
 &&  \sigma^R_{13}(t,z;\vec q)=\sin|\Omega|t\;
 \sigma^W_{12}(T_W,z;\vec q)- \nn\\
 &&-g^2N\int_0^t dt^\prime \cos|\Omega|(t-t^\prime)\; \int_0^{t^\prime} dt^{\prime\prime}
\sin|\Omega|t^{\prime\prime} \int_0^z dz^\prime\sigma^W_{12}(T_W,z^\prime;\vec q)D(t^\prime-t^{\prime\prime}, z-z^\prime),\L{37}\\
 &&\sigma^R_{12}(t,z;\vec q)=\cos|\Omega|t\;
 \sigma^W_{12}(T,z;\vec q)+\nn\\
 &&+g^2N\int_0^t dt^\prime \sin|\Omega|(t-t^\prime)\int_0^{t^\prime} dt^{\prime\prime}\;
\sin|\Omega|t^{\prime\prime} \int_0^z dz^\prime\sigma^W_{12}(T_W,z^\prime;\vec q)D(t^\prime-t^{\prime\prime},
z-z^\prime),\L{38}
\EY
\end{widetext}
where the ground state coherence at the beginning of the process
$\sigma^W_{12}(T_W,z^\prime,\vec q)$ contains the stored signal
field. This value coincides with equation (\ref{21}) at $t=T_W$.

Solutions for the backward read-out are similar except for the spatial argument in the coherence $\sigma^W_{12}(T_W,z^\prime;\vec q)$ where
$z^\prime$ has to be changed to $L-z^\prime$. Substituting the coherence $\sigma_{12}$ from Eq.(\ref{21}) to Eq.(\ref{36}) and taking into account
the diffraction factors (\ref{16}) one can derive the signal field at the output of the memory cell for the forward read-out of the form
 \BY
&&  a_{forward}^R(\tilde t,\tilde L;\vec q)=\frac{1}{2}a_{in}(\vec q)\;e^{-iq^2L/(2k_s)}\int_0^{\tilde L} d\tilde
z^\prime\nn\\
&&\[\sin\tilde t+\int_0^{\tilde t} d\tilde t^\prime \sin\tilde t^\prime
 \; \tilde D(\tilde t-\tilde t^\prime, \tilde L-\tilde z^\prime)\]\times\L{4.4}\\
 &&\[1-\cos\tilde
T_W+\int_0^{\tilde T_W} d\tilde t^{\prime\prime} \[1-\cos(\tilde T_W-\tilde t^{\prime\prime})\]
 \; \tilde D(\tilde t^{\prime\prime}, \tilde z^\prime)\]\nn
\EY
and correspondingly for the backward read-out:

%
 \BY
&&  a_{back}^R(\tilde t,\tilde L;\vec q)=\frac{1}{2}a_{in}(\vec q)\int_0^{\tilde L} d\tilde z^\prime
e^{-iq^2z^\prime/k_s}\nn\\
&&\[\sin\tilde t+\int_0^{\tilde t} d\tilde t^\prime \sin\tilde t^\prime
 \; \tilde D(\tilde t-\tilde t^\prime, \tilde z^\prime)\]\times\L{4.5}\\
 &&\[1-\cos\tilde
T_W+\int_0^{\tilde T_W} d\tilde t^{\prime\prime} \[1-\cos(\tilde T_W-\tilde t^{\prime\prime})\]
 \; \tilde D(\tilde t^{\prime\prime}, \tilde z^\prime)\].\nn
\EY
%
%
These equations are written using the dimensionless time and space coordinates except for the exponential diffraction factor where the regular $z$
coordinate is kept. One can see that the situation is very different for forward and backward retrieval. For forward read-out the diffraction
introduces a common factor which can easily be compensated with an appropriate lens. Thus, in this case, although diffraction takes place separately
in
 writing and reading, the effects of diffraction due
 to the writing process can be perfectly compensated under the
read-out.

A dramatically different situation takes place for the backward read-out. Because the diffraction factor is under the integral, the diffraction are
able to modify seriously the initial transverse distribution. In contrast to the first case, the diffraction processes in the writing and reading
channels do not compensate each other but on the contrary add their effects. From Eqs. (\ref{4.4})-(\ref{4.5}), it can be seen that under the
geometrical conditions $q^2L/k_s\ll1$, diffraction is not detrimental (see also
 below). Here we shall consider only this case and we show in Fig.~8 the time dependence of the intensity of
the read-out signal.

\begin{figure*}
\includegraphics[height=5cm]{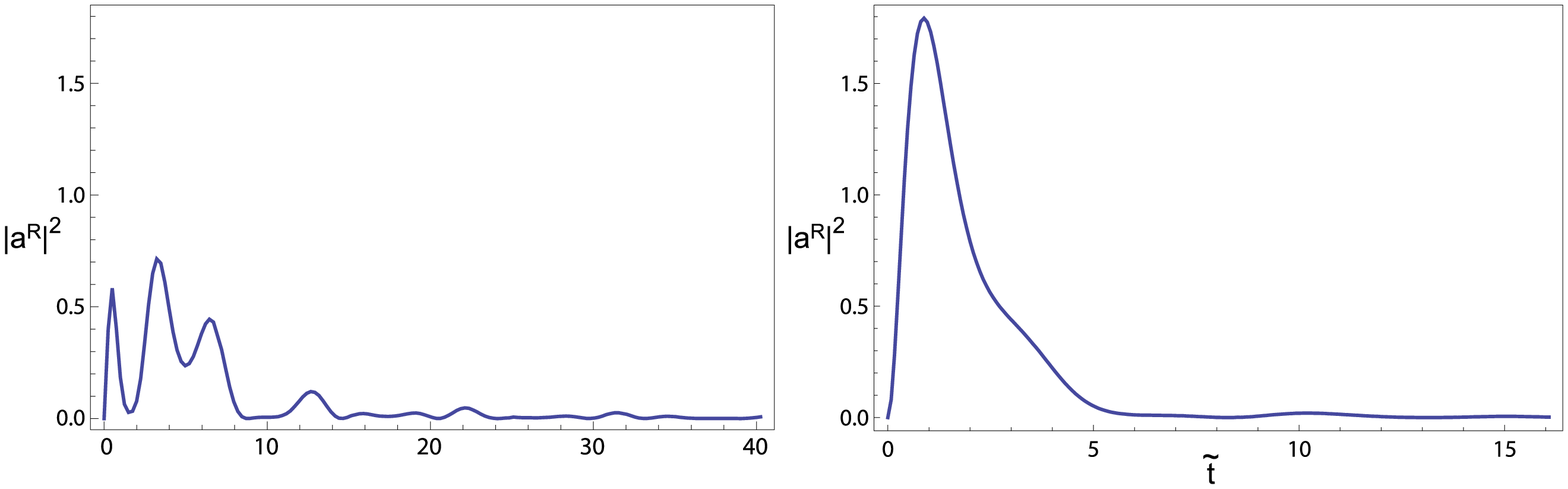}
 \caption{Reading process : field intensity $|a^R(\tilde t,\tilde L)|^2$ at the output of the medium for $\tilde L=10.3$
 and $\tilde T_W=4.2$ for (a) forward and (b) backward propagating retrieval.}
 \label{fig9}
\end{figure*}

\subsection{Efficiency of the memory process}

Fig. \ref{fig9} shows the intensity of the re-emitted normalized signal field $a^R(\tilde t,\tilde L)=a^R(\tilde t,\tilde L,\vec q)/a_{in}(\vec q)$
at the output of the medium as a function of the reading time for forward and backward retrieval for an optical depth $\tilde L=10.3$ and a writing
time duration $\tilde T_W=4.2$. It can be seen that the shape of the restored pulse is quite different from the shape of the input signal pulse.
However, we will not focus on restoring the pulse shape, but rather, on the read-out of an image corresponding to an ensemble of the transverse modes
of the signal field, which has been stored in the atomic medium as a quantum hologram. To do this, the photon number in each of the reconstructed
transverse modes must be the same as the photon number in the corresponding modes of the original field. To estimate the storage efficiency, we
introduce the parameter $ Eff (\vec q) $, defined as the ratio of the integrated intensity of the recovered pulse in mode $ \vec q $ to the
integrated intensity of the same transverse mode in the input pulse :
\BE Eff(\vec q)=\frac{\int_0^{\tilde T_R}|a^R(\tilde t,\tilde z,\vec q)|^2 d \tilde t}{\int_0^{\tilde T_W}| a_{in}(\tilde t,\vec q)|^2 d \tilde
t\L{Eff}}\times100\%
\EE

where $ T_R $ is the reading time duration. It can be seen from Fig. \ref{fig9} that for an efficient recovery, for forward retrieval, the reading
time duration should substantially exceed the writing time duration. For example, in Fig. \ref{fig9}a, the retrieval efficiency after a reading time
equal to the writing time $ T_W $ is only 36\%. For backward retrieval, the retrieval is much more efficient. In Fig. \ref{fig9}b the efficiency
reaches 80\% for a reading  time $ T_R=T_W $.

Losses come from both the writing and the reading processes. With parameters $\tilde L=10.3$ and $\tilde T_W=4.2$ Fig. \ref{fig4} shows that the
losses in the
 writing process are 5.1 \%. Then, the best efficiency of  the whole process, including writing and
 read-out cannot exceed  $ Eff = 94.9 \% $.

For the reading process, a numerical calculation shows that if the reading time is $ 10 $ times the writing time $ T_R = 10T_W $, the overall
efficiency is $ Eff = 77\% $ for forward retrieval. We could choose an even longer reading duration, but we must remain in the approximation that $
T_R \ll\gamma^{-1} $. The backward retrieval is faster and the same calculation yields an overall efficiency of 83.8\% for $T_R = 3T_W $.

In the same way as for writing, we observe oscillations in the read-out field in forward direction (Fig.
\ref{fig9}a). This indicates that the reading process is accompanied by a rewriting process in the medium, before
emission. In the case of backward retrieval, the situation is much more favorable and the curve does not show any
oscillation (Fig. \ref{fig9}b).

The read-out as well as the writing is coupled with a displacement of the coherence along the medium. From the results obtained in the previous
section it can be seen that for not too long writing pulses, the storage in the atomic coherence is mainly localized in the input layers of the
medium. Then, if the medium is long enough, the forward retrieval involves reading the signal and writing it again on successive layers in the
medium, which eventually "pushes" the read-out signal out of the medium. On the contrary, in the backward retrieval there is less "pushing", since a
large coherence is concentrated at the exit face.

\subsection{Restoring the transverse profile }

Let us now examine in more details the information about the transverse profile of the field in the reconstructed pulse. Usually, the quantum memory
of the pulse implies its longitudinal profile storage. A lot of works devoted to this problem (for example,
\cite{Nunn},\cite{Gorshkov2},\cite{Gorshkov4}) have focused on the determination of eigenfunctions of the system, we will not deal with this issue.
Our goal is to restore the transverse profile of the signal field. To achieve a quantum memory we need to ensure high efficiency of reconstruction
for each transverse mode. In the previous section, we showed that each of the transverse modes can be recovered with equal efficiency, with a value
that depends on the choice of experimental parameters and that can be high enough. However for the quantum storage of the transverse profile we
should have not only a high efficiency for each component but also an insensibility to the diffraction phenomena.

In the previous section we showed that the reading conditions are different depending on whether reading is a direct or reverse process. However,
this apparent advantage of backward reading is associated with a limitation due to diffraction phenomena. The essential difference between the
forward and backward reading is that in the first case, the diffraction distortions that occur during storage are almost entirely compensated under
subsequent reading. What is left uncompensated can be easily cleaned, for example, by setting additional lens at the output of the cell.

 In the second case, the diffraction distortions that occur sequentially during recording and reading,
 add to each other, and they have to be minimized independently.
 It can be seen that this leads to a significant limitation for an important memory
 parameter, which is the number of modes  or the grain of the image that can be
 efficiently stored in the memory cell. In principle, this number of modes is
 determined by the ratio between the transverse size of the memory cell $S$ and the grain size $d$ of the image at
 the entrance to the cell and is equal to $N = S / d ^ 2$. Minimizing the diffraction means
 that at the output of the memory cell of thickness $L$ the grain size $D = L \lambda / d$ should coincide with the input grain
 size $D = d$. This means that the number of stored modes N can not exceed the Fresnel number $F_N = S / (\lambda L)$.

So our numerical analysis of backward reading that does not take diffraction into account can be considered as justified, provided that the number of
stored modes is below the Fresnel number $N \leq F_N$.

Similarly, we can estimate the number N for forward reading. In this case, as already mentioned, we can generally ignore the diffraction image
distortion associated with the intersection of various grains at the cell output. However, we should not allow excessive beam divergence and, as a
result of this divergence, light losses due to leakage of the field through the side walls of the cell. To avoid these losses, the output grain size
should not exceed the transverse size of the memory cell $D\leq\sqrt S  $. It is easy to see that the number of stored modes should not exceed the
square of the Fresnel number $N \leq F_N ^ 2$, in agreement with Ref.\cite{Denis2}. We can thus predict a large storage capacity for this
$\Lambda$-based memory scheme.

\section{Conclusion}

We have presented a full calculation of the writing and read-out processes of a short pulse in an atomic medium. Our approach relies on a full
treatment of light matter interaction as the writing or reading pulses propagate through the atomic medium. This allows to obtain detailed
predictions on the behavior of the atomic ground state coherence, which stores the quantum information and on the re-emitted field. As a result,
precise values of the storage efficiency and of the memory capacity for multimode storage are given, together with a detailed optimization procedure
for the memory operation. We have also shown that if backward retrieval provides a better efficiency for a single mode, in the case of spatial
multimode storage it tends to limit the number of stored modes to a lower value than the forward retrieval. With the experimental developments of
short pulse storage and fast memory registers, our model provides a reliable method to optimize the experimental conditions.




\section{Acknowledgements}

We thank Ivan Sokolov and Denis Vasilyev for the interesting
discussion and remarks. The study was performed within the
framework of the Russian-French Cooperation Program "Lasers and
Advanced Optical Information Technologies", of the European
Project HIDEAS (grant No. 221906) and of the Ile de France
programme IFRAF. O.M. acknowledges the support of IFRAF. A.B. is a
member of the Institut Universitaire de France. The study was also
supported by RFBR (grant No. 08-02-92504).

\appendix


\section{Main equations and general solutions\L{A}}

 The light matter interaction Hamiltonian for our
problem can be written as
%
\BY
\hat V=\int dz \;d^2\rho \[ ig \right.&&\(\hat a(z,\vec\rho,t)e^{\ds i k_s z-i\Delta t}\hat\sigma_{31}(z,\vec\rho,t)-\nn\right.\\
&&\left.\hat a^\dag(z,\vec\rho,t)e^{\ds -i k_s z+i\Delta t}\hat\sigma_{13}(z,\vec\rho,t)\)+\nn\\
&& i\Omega (t)e^{\ds i k_d z-i\Delta t}\hat\sigma_{32}(z,\vec\rho,t)-\nn\\
&&\left. i\Omega^\ast(t)e^{\ds -i k_d z+i\Delta t}\hat\sigma_{23}(z,\vec\rho,t)\],\nn\\
&&\qquad \vec\rho= \vec\rho(x,y),\L{A1} \EY
%
where $k_s$, $k_d$ are wave vectors of the signal, driving waves, and the value $\Delta $ determines the two-photon resonance
\BY
&&\Delta=\omega_s-\omega_{13}=\omega_d-\omega_{23}.
\EY

We perform the substitutions
\BY &&\hat \sigma_{13} \to e^{\ds i k_s z-i\Delta t} \hat \sigma_{13},\qquad\hat \sigma_{23} \to e^{\ds i k_d
z-i\Delta t} \hat \sigma_{23},\nn\\
&&\hat \sigma_{12} \to e^{\ds -i (k_d-k_s) z} \hat \sigma_{12}.
 \EY
Then the Heisenberg evolution equation for the atom field system can
be written as
\BY
&& \( \frac{\partial}{\partial t}+c\frac{\partial}{\partial z}-\frac{ic}{2 k_s}\Delta_\perp \)
\hat a=-c g \hat\sigma_{13},\L{A3}\\
&& \frac{\partial}{\partial t}{\hat \sigma}_{13}= -i\Delta{\hat \sigma}_{13}+\Omega \hat\sigma_{12} + g \hat a (\hat N_1-\hat N_3),\L{A4}\\
&& \frac{\partial}{\partial t}{\hat \sigma}_{12}= - \Omega^\ast \hat\sigma_{13} - g \hat a
\hat\sigma_{32},\L{A5}\\
&& \frac{\partial}{\partial t}{\hat \sigma}_{32}= i\Delta{\hat \sigma}_{32}- \Omega^\ast (\hat N_3-\hat N_2) + g \hat
a^\dag
\hat\sigma_{12},\L{A6}\\
&&\frac{\partial}{\partial t}{\hat N}_1= - g \hat a \hat\sigma_{31} -
g \hat a^\dag  \hat\sigma_{13},\\
&& \frac{\partial}{\partial t}{\hat N}_2= - \Omega \hat\sigma_{32} -
\Omega^\ast \hat\sigma_{23},\\
&& \frac{\partial}{\partial t}{\hat N}_3=-\frac{\partial}{\partial t}{\hat N}_1-\frac{\partial}{\partial t}{\hat
N}_2.\L{A9}
\EY
where the transverse Laplace operator reads
\BY
&&\Delta_\perp=\frac{\partial^2}{\partial x^2}+\frac{\partial^2}{\partial y^2}.
\EY

Here we do not take into account a spontaneous emission with rate $\gamma$  on the transition $|3\rangle\to|1\rangle$. This is coupled with a
requirement that in our model the field pulses are very short such that the spontaneous emission has no time to introduce something to the atomic
state.

By ignoring in Eqs.~(\ref{A3})-(\ref{A9}) the spontaneous relaxation we nevertheless survive the frequency detuning $\Delta$. The condition
$\Delta\gg\gamma$ usually means that we want to treat the Raman process. However for the short pulses this condition do not ensure Raman process,
here Raman interaction is realized only in the case of stronger inequality: $\Delta\gg T^{-1}$. In this article we consider the opposite condition
$\Delta\ll T^{-1}$, when we are under obligatory have to neglect by $\Delta$ in Eqs.~(\ref{A4}) and (\ref{A6}).

From Eqs.~(\ref{A3})-(\ref{A9}) one can obtain that the operator
\BY
&&\hat S(t)=\int d^3r\(\hat n(\vec r,t)/c+\hat N_2(\vec r,t)+\hat N_3(\vec r,t)\)\;\;\;\;\;\;
\EY
survives in time. Here $\hat n(\vec r,t)=\hat a^\dag(\vec r,t)\hat a(\vec r,t)$ is the photon number operator per second per unit area. Physically
this result seems quite natural.

We assume the following conditions. All the $N$ atoms are initially in the state $|1\rangle$ and we assume that the signal field is much weaker than
the driving field $|\Omega|^2~\gg~g^2~\langle\hat a^\dag\hat a\rangle$. Then most of the atoms remain in the ground state $|1\rangle$ and in
Eq.~(\ref{A4}) in approximation of the quasi-homogeneous medium we have a right to make a change $(\hat N_1-\hat N_3)\to N$ and also we can neglect
the term proportional to $\hat\sigma_{32}$ in Eq.~(\ref{A5}). Thus we obtained a reduced closed system of equations for interesting values
\BY
&& \(\frac{1}{c}\frac{\partial}{\partial t}+\frac{\partial}{\partial z}-\frac{i}{2 k_s}\Delta_\perp \) \hat
a(z,\vec\rho,t)=\nn\\
&&- g\;
\hat\sigma_{13}(z,\vec\rho,t),\L{A19}\\
 &&\frac{\partial}{\partial t}{\hat\sigma}_{13}(z,\vec\rho,t)=  gN \hat a(z,\vec\rho,t)+\Omega
 \hat\sigma_{12}(z,\vec\rho,t),\;\;\;\;\\
 && \frac{\partial}{\partial t}{\hat\sigma}_{12}(z,\vec\rho,t)= - \Omega^\ast\hat\sigma_{13}(z,\vec\rho,t).\L{}
\EY
We neglect the time delay linked to the pulse propagation in the atomic medium. This means if we have long enough pulses, such that $L/c\ll T$ ($L$
is thickness of the medium and $T$ is the pulse duration), we can neglect the time interval between the time at which the front part of the pulse
enters the medium and the time at which the front part leaves it. Formally this means we can neglect the time derivative in Eq.~(\ref{A19}).

Let us take the Fourier transform  and next the Laplace transform of
the equations according to the relations
\BY
&& \hat F(z,t;\vec q)=\frac{1}{2\pi}\int \hat F(z,\vec\rho,t)e^{\ds -i \vec q\vec\rho}\;d^2\rho\L{A22},\\
&&\hat F_s(z;\vec q)=\int_0^\infty dt F(z,t;\vec q)\;e^{-st}.\L{A23}
 \EY
In the Fourier domain the equations read
\BY
&& \frac{\partial}{\partial z} \hat a(z,t;\vec q)=- g \;\hat\sigma_{13}(z,t;\vec q),\L{A24}\\
 &&\frac{\partial}{\partial t}{\hat\sigma}_{13}(z,t;\vec q)=  gN \hat a(z,t;\vec q)+\Omega \hat\sigma_{12}
 (z,t;\vec q),\;\;\;\;\L{A25}\\
 && \frac{\partial}{\partial t}{\hat\sigma}_{12}(z,t;\vec q)= - \Omega^\ast\hat\sigma_{13}(z,t;\vec q)\L{A26}.
\EY
where we have made the changes
\BY
 && \hat a(z,t;\vec q)\to\hat a(z,t;\vec q) e^{\ds -i q^2 z/(2k_s)},\nn\\
 &&
 \hat \sigma_{mn}(z,t;\vec q)\to\hat \sigma_{mn}(z,t;\vec q) \;e^{\ds -i q^2 z/(2k_s)}.\;\;\;\;\L{A27}
 \EY
In the Laplace domain the equations are written in the form
\BY
&& \frac{d \hat a_s(z;\vec q)}{d z} =- g \hat \sigma_{13, s}(z;\vec q)\L{A28}\\
 &&-\hat \sigma_{13}(z,0;\vec q)+s\hat \sigma_{13, s}(z;\vec q)= gN  \hat a_s(z;\vec q)+ \nn\\
 &&
 \Omega \hat \sigma_{12, s}(z;\vec q),\\
 && -\hat \sigma_{12}(z,0;\vec q)+s\hat \sigma_{12, s}(z;\vec q)=\nn\\
 && - \Omega^\ast\hat \sigma_{13, s}(z;\vec q) \L{A30}.
\EY
Eliminating the coherences $\sigma_{13,s}$ and $\sigma_{12,s}$ we
obtain a differential equation for the Laplace field amplitude as :
 \BY
&&\frac{d\hat a_s(z;\vec q)}{dz}=-\gamma_s\hat a_s(z;\vec q)-g\hat A_s(z;\vec q),\L{A31}
\EY
where
 \BY
&&\gamma_s=g^2N\frac{s}{s^2+|\Omega|^2},\\
&& \hat A_s(z;\vec q)=\frac{1}{s^2+|\Omega|^2}\[\Omega\hat \sigma_{12}(0,z;\vec q)+s\hat \sigma_{13}(0,z;\vec
q)\].\nn
\EY
The solution of equation (\ref{A31}) reads
 \BY
&&\hat a_s(z;\vec q)=\hat a_s(0;\vec q)e^{\ds-\gamma_sz}-\nn\\
&&g\int_0^zdz^\prime \hat A_s(z^\prime;\vec q)e^{\ds-\gamma_s(z-z^\prime)}.\L{A33}
\EY
Taking the inverse Laplace transform we obtain the field amplitude
expressed as a function of the initial conditions
\BY
&& \hat a(t,z;\vec q)=\int_0^t dt^\prime \hat a_{in}(t-t^\prime;\vec q)\; D(t^\prime, z) -\nn\\
&&g\int_0^t dt^\prime\int_0^z dz^\prime \hat A(t-t^\prime, z-z^\prime;\vec q)\; D(t^\prime, z^\prime).\;\;\;\;\L{A34}
\EY
The kernel $D(t,z)$ is expressed from the Bessel's functions and the
function $A(t, z;\vec q)$  reads

\begin{widetext}
 \BY
 &&D(z,t)=\delta(t)-\cos|\Omega|t\;\sqrt{\frac{2g^2Nz}{t}}J_1\(\sqrt{2g^2Nzt}\)+\nn\\
 &&\nn\\
 &&+\frac{1}{2}g^2Nz\int_0^tdt^\prime
\[ \frac{1}{\sqrt{t^\prime}}e^{\ds -i|\Omega|t^\prime}J_1\(\sqrt{2g^2Nzt^\prime}\)\]\[ \frac{1}{\sqrt{t-t^\prime}}
e^{\ds i|\Omega|(t-t^\prime)} J_1\(\sqrt{2g^2Nz(t-t^\prime)}\)\],\L{A35}
\EY
\BY
 && \hat A(t,z;\vec q)=\cos|\Omega|t\;\hat \sigma_{13}(0,z;\vec q)+\sin|\Omega|t\;\hat \sigma_{12}(0,z;\vec q).\L{A36}
\EY

From Eqs.~(\ref{A25}) and (\ref{A26}) we also get explicit
solutions that read
\BY
&& \hat \sigma_{13}(t,z;\vec q)= gN\int_0^t dt^\prime \cos|\Omega|(t-t^\prime)\; \hat a(t^\prime,z;\vec q)+\hat
A(t,z;\vec
q),\L{A37}\\
 &&\hat\sigma_{12}(t, z;\vec q)=e^{\ds -i\varphi_\Omega}\[-gN\int_0^t dt^\prime \sin|\Omega|(t-t^\prime)
 \; \hat a(t^\prime,z;\vec q)+ \hat B(z,t;\vec q)\]\L{A38},
\EY
where
\BY
 &&\hat B(z,t;\vec q)=-\sin|\Omega|t\;\hat \sigma_{13}(0,z;\vec q)+\cos|\Omega|t\;\hat \sigma_{12}(0,z;\vec q).\L{A39}
 \EY
Thus we have obtained the solutions of the system of the main equations under the arbitrary initial conditions.
\end{widetext}

\end{document}